\documentclass{webofc}
\usepackage[varg]{txfonts}   
\usepackage{graphicx}
\usepackage{amsmath}
\usepackage{amssymb}
 \usepackage{url} 
 \newcommand{\open}{\sphericalangle}
\begin{document}
\title{The Tensor and the Scalar Charges of the Nucleon from
Hadron Phenomenology}
%
%

\author{\firstname{A.} \lastname{Courtoy}\inst{1}\fnsep\thanks{\email{aurore@fisica.unam.mx}}         
}

\institute{Instituto de F\'isica, Universidad Nacional Aut\'onoma de M\'exico\\
Apartado Postal 20-364, Ciudad de M\'exico 01000, M\'exico
          }

\abstract{%
We discuss the impact of the determination of the nucleon tensor charge on searches for physics Beyond the Standard Model. We also comment on the future extraction of the subleading-twist PDF $e(x)$ from Jefferson Lab soon-to-be-released Beam Spin Asymmetry data as well as from the expected data of CLAS12, as the latter is related to the scalar charge.
These analyses are possible through the phenomenology of Dihadron Fragmentation Functions related processes, which we  report on here as well. 
}
\maketitle
\section{Introduction}
\vspace{.2cm}

In these proceedings, we discuss some properties of the nucleons that are of interest for both hadronic physics and physics Beyond the Standard Model (BSM): their tensor and the scalar charges. Both can, in principle, be accessed through the phenomeology of Parton Distribution Functions. The latter are related to the charges via specific sum rules.
Due to their chiral-odd nature, the respective parton distributions are involved in processes together with another non-perturbative chiral-odd partner. In semi-inclusive DIS, that partner is a fragmentation function. So to examine collinear/1-dimensional parton distributions, we will  consider  Dihadron Fragmentation Functions (DiFF), which encode information about the fragmentation process of a quark into a hadron pair (plus something else, undetected). Fragmentation has been hardly studied through models for pion pairs, the main knowledge about those DiFFs  being for now fits from data.

The rich phenomenology associated to dihadron fragmentation compares, though with less statistics, to the well-known Semi-Inclusive Deep Inelastic Scattering (SIDIS) in which a single hadron is produced in the current fragmentation region. Once the distribution functions are extracted through SIDIS asymmetries, their $x$-dependence can be integrated to obtain the charges, as dictated by the sum rules. 
We further discuss the future of dihadron data and their impact on physics BSM. 

\section{The Dihadron Way to Distribution Functions}

\vspace{.2cm}

The phenomenological knowledge on DiFFs is based on $e^+-e^-$ annihilation at Belle. The unpolarized DiFF has been extracted by fitting the pion pair distribution simulated by a Monte Carlo event generator~\cite{Courtoy:2012ry,Radici:2015mwa} to compensate for the absence of data on multiplicities at that time. The resulting error on the parametrization of the unpolarized DiFF is obviously small. On the other hand, once the unpolarized DiFF parameterized, the chiral-odd DiFF has been extracted from the Artru--Collins asymmetry at Belle. The factorization of the process
$e^+ e^- \to (\pi^+ \pi^-)_{\mbox{\tiny jet}} (\pi^+ \pi^-)_{\overline{\mbox{\tiny jet}}} X$ is valid in the kinematical regime $P_h^2 = M_h^2 \ll Q^2=-q^2 \geq 0$ and $q = k - k'$ is the space-like momentum transferred, with the total momentum $P_h = P_1 + P_2$.

\begin{figure}
\centering
\includegraphics[width=8cm]{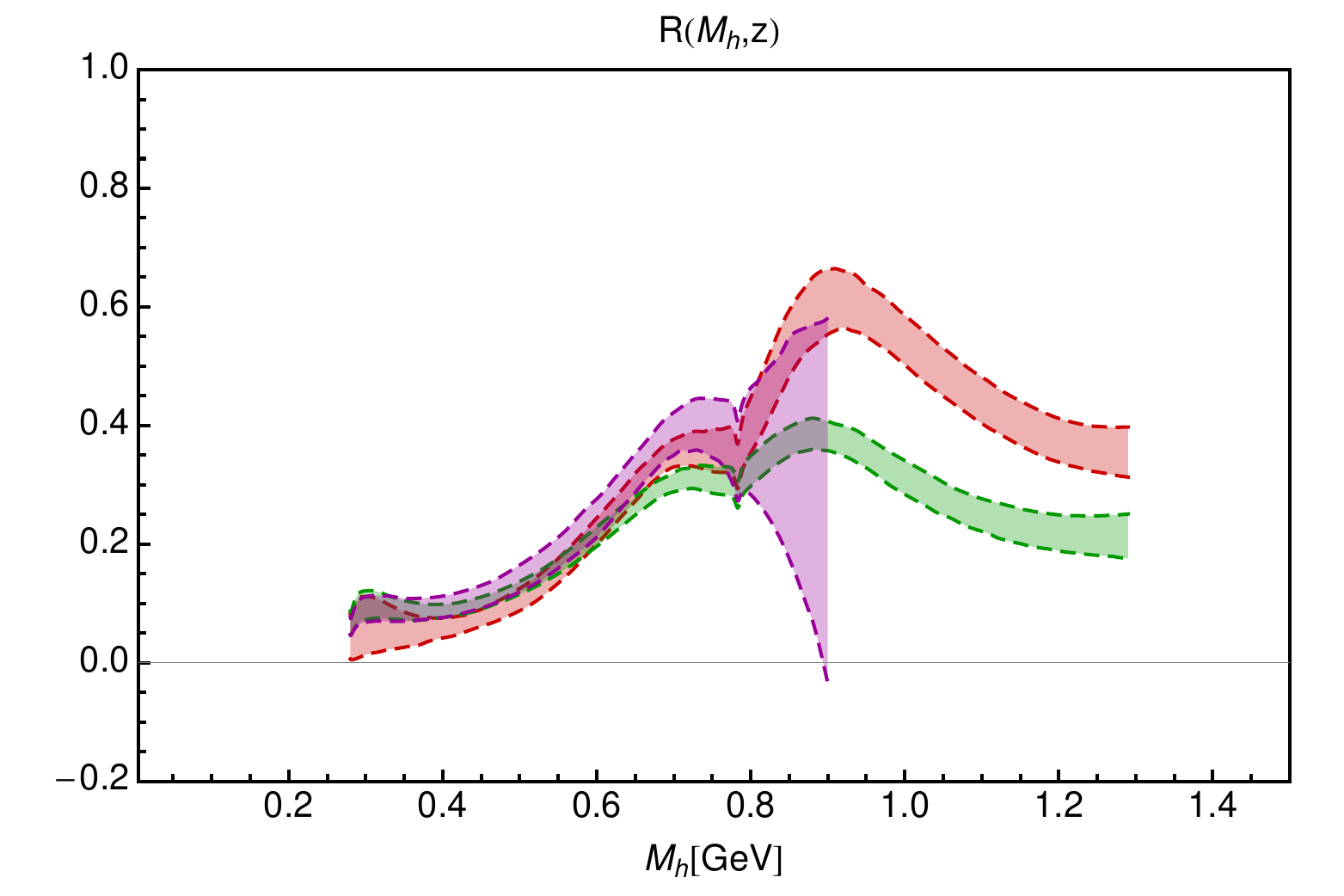} 
\caption{The ratio $R(z, M_h)$ as a function of $M_h$ at $Q_0^2=1$ GeV$^2$ for three different $z=0.25$ (shortest band), $z=0.45$ (lower band at $M_h \sim 1.2$ GeV), and $z=0.65$ (upper band at $M_h \sim 1.2$ GeV), with the value $\alpha_s (M_Z^2) = 0.139$ used in the QCD evolution equations. } 
\label{fig:H1Mh}
\end{figure}

 Belle data for the Artru--Collins asymmetry led to the extraction ---and, hence, parameterization--- of the leading-twist DiFFs~\cite{Courtoy:2012ry,Radici:2015mwa}, illustrated in Fig.~\ref{fig:H1Mh} through their ratio
\begin{equation}
R(z, M_h) = \frac{|\bf{R}|}{M_h}\, \frac{H_1^{\open\, u} (z, M_h; Q_0^2)}{D_1^u (z, M_h; Q_0^2)} \; ,
\label{e:R}
\end{equation}
 the relevant variables being the pion pair invariant mass,  $M_h$, and the sum of the fractional energies carried by the two final hadrons, $z$. 
 
 While the first fitting approach was based on the usual Hessian statistics with a variation of the chisquare equal to unity~\cite{Courtoy:2012ry}, the newest version of the fitting procedure incorporates a Neural Network preparation, {\it i.e.} the normal generation of data replicas within the ---$1\sigma$---  experimental errors~\cite{Radici:2015mwa}. This second technique insures the proper treatment of errors outside the data range, thus avoiding the vanishing of the errors at the endpoints, where they should be maximal. We have noticed that both fitting techniques, still based on adhoc functional forms, agree with each other as far as the chiral-odd DiFF is concerned.

%



Combining the parametrization obtained for DiFFs to HERMES and COMPASS data on  dihadron SIDIS off transversly polarized target allowed for the extraction of the last leading-twist PDF, {\it i.e. the transversity PDF}~\cite{Radici:2015mwa,Bacchetta:2012ty,Bacchetta:2011ip}. 
The structure functions can be written as products of distribution functions and 
DiFFs:
\begin{align} 
F_{UU} & = x \sum_q e_q^2\, f_1^q(x; Q^2)\, D_1^q\bigl(z,\cos \theta, M_h; Q^2\bigr) \; , 
\label{e:FUU} \\
F_{UT}^{\sin (\phi_R +\phi_S)} &=  \frac{|\bf R| \sin \theta}{M_h}\, x\, 
\sum_q e_q^2\,  h_1^q(x; Q^2)\,H_1^{\open\, q}\bigl(z,\cos \theta, M_h; Q^2\bigr) \; , 
\label{e:FUT}
\end{align}
where $e_q$ is the fractional charge of a parton with flavor $q$ and the distribution functions $f_1^q(x)$ and $h_1^q(x)$ are, respectively, the unpolarized and the transversity PDFs.

While the fit of the transversity is an achievement by itself, though affected of a large uncertainty outside the data kinematical range, its first Mellin moment is also of great interest. The tensor charge is obtained by integrating the transversity PDF over the physical support in $x$,
\begin{eqnarray}
\delta q_v (Q^2) &= &\int_0^1 dx \, h_1^{q_v} (x, Q^2) \; . 
\label{e:tensch}
\end{eqnarray}
The error we mentioned above, consequence of the extrapolation of the PDF outside the data range, is the main source of uncertainty on the determination of the tensor charge. For example, going from a truncated integration over $x$ to a full integration increases the central value up to a factor of 2 and the error up to a factor of 3 for the up quark and factors of 4 and 7 for the down quark tensor charge respectively, as can be seen in Table 3 of Ref.~\cite{Radici:2015mwa}. Another source of error consists in the choice of the functional form for the transversity PDF. This problem has been taken care of, in a first step, considering three different functional forms,  each with a growing number of free parameters. The value for the isovector tensor charge $g_T = \delta u_v - \delta d_v$   for the functional form related to the so-called {\it flexible scenario} with $\alpha_s (M_Z^2)=0.125$ is, at $1\sigma$,
\begin{eqnarray}
g_T = 0.81 \pm 0.44\quad \mbox{ at} \quad Q^2=4 \, \mbox{GeV}^2\quad.
\label{eq:gtpavia}
\end{eqnarray}
 It is in agreement with lattice determinations as well as with the other extractions from hadronic phenomenology \cite{Anselmino:2013vqa,Goldstein:2014aja,Kang:2015msa,Ye:2016prn}, though the absolute value is slightly smaller than the lattice's. Due to its non-perturbative nature, the nucleon structure can only be unveiled using complementary methods such as effecive field theories, lattice calculations, models for the nucleon structure, Schwinger-Dyson based techniques and phenomenological extraction of observables from data. As such, the lattice field theory and the data based determinations play a similar, yet uncorrelated, role. 

With the future JLab data, we will analyze the 4-dimensional binning spanning a wide kinematical range in $(z, M_h, x, Q^2)$, for both proton and deuteron targets, allowing for a flavor separation of the valence distributions. 
The development expected from such a measurement and statistical analysis will qualitatively improve the present estimates for a number of processes, {\it e.g.} we expect about $10\%$ improvement in the errorbars with CLAS12 and SoLID data~\cite{Courtoy:2015haa}. 
We would like to restate that DiFFs are universal and can, as such, be transported to, {\it e.g.}, SIDIS or proton-proton asymmetries.
This characteristic has recently been studied and reinforced through a confirmation on the role of transversity in proton-proton collisions at RHIC~\cite{Radici:2016lam}.
\\

Though in a more elusive way, the scalar charge is related to a twist-3 PDF, $e(x)$.  QCD equations of motion allow to decompose the chiral-odd twist-3 distributions into three terms%
\begin{eqnarray}
e^q(x)&=&e_{\mbox{\tiny{loc}}}^q(x)+e_{\mbox{\tiny{tw-3}}}^q(x)+e_{\mbox{\tiny{mass}}}^q(x)\quad.
\end{eqnarray}
The first term comes from the local operator and is exactly the contribution related to scalar charge,
\begin{eqnarray}
e_{\mbox{\tiny{loc}}}^q(x)&=&\frac{1}{2M}\, \int \frac{d\lambda}{2\pi}\, e^{i\lambda x}\langle P| \bar{\psi}_q(0)\psi_q(0)|P\rangle\quad,\nonumber\\
&=&\frac{\delta(x)}{2M}\langle P| \bar{\psi}_q(0)\psi_q(0)|P\rangle\quad;
\end{eqnarray}
the second term is a genuine twist-3 contribution, {\it i.e.} pure quark-gluon interaction term ; while the last term is related to the current quark mass.

The sum rule related the chiral-odd twist-3 PDF to the scalar charge is similar to Eq.~(\ref{e:tensch}). Only the {\it local} term contributes to the first Mellin moment, {\it i.e.} the scalar charge is nothing else than $e(x=0)$.  That delta-function singularity follows from chiral symmetry and the existence of non-vanishing quark condensate~\cite{Wakamatsu:2003uu}.

The chiral-odd twist-3 PDF is  accessible in single-~\cite{Efremov:2002ut} and di-hadron SIDIS.  In the latter case, the chiral-odd partner of $e(x)$ is the chiral-odd DiFF:
\begin{align} 
F_{LU}^{\sin\phi_R}   = -\frac{|{\bf R}| \sin \theta}{Q}\,x\,\sum_q e_q^2\;\biggl[
    \frac{M}{m_{hh}}\,x\, e^q(x; Q^2)\, H_1^{\open\, q}\bigl(z,\cos \theta, M_h; Q^2\bigr)
    +\frac{f_1^q(x; Q^2)}{z}\,\widetilde{G}^{\open\, q}\bigl(z,\cos \theta, M_h; Q^2\bigr)\biggr]\,,
    \label{e:FLU}
\end{align}
where the second term of the structure function contains the direct product of the unpolarized PDF and an --unknown-- twist-3 DiFF.

\begin{figure}
\centering
\includegraphics[width=8cm]{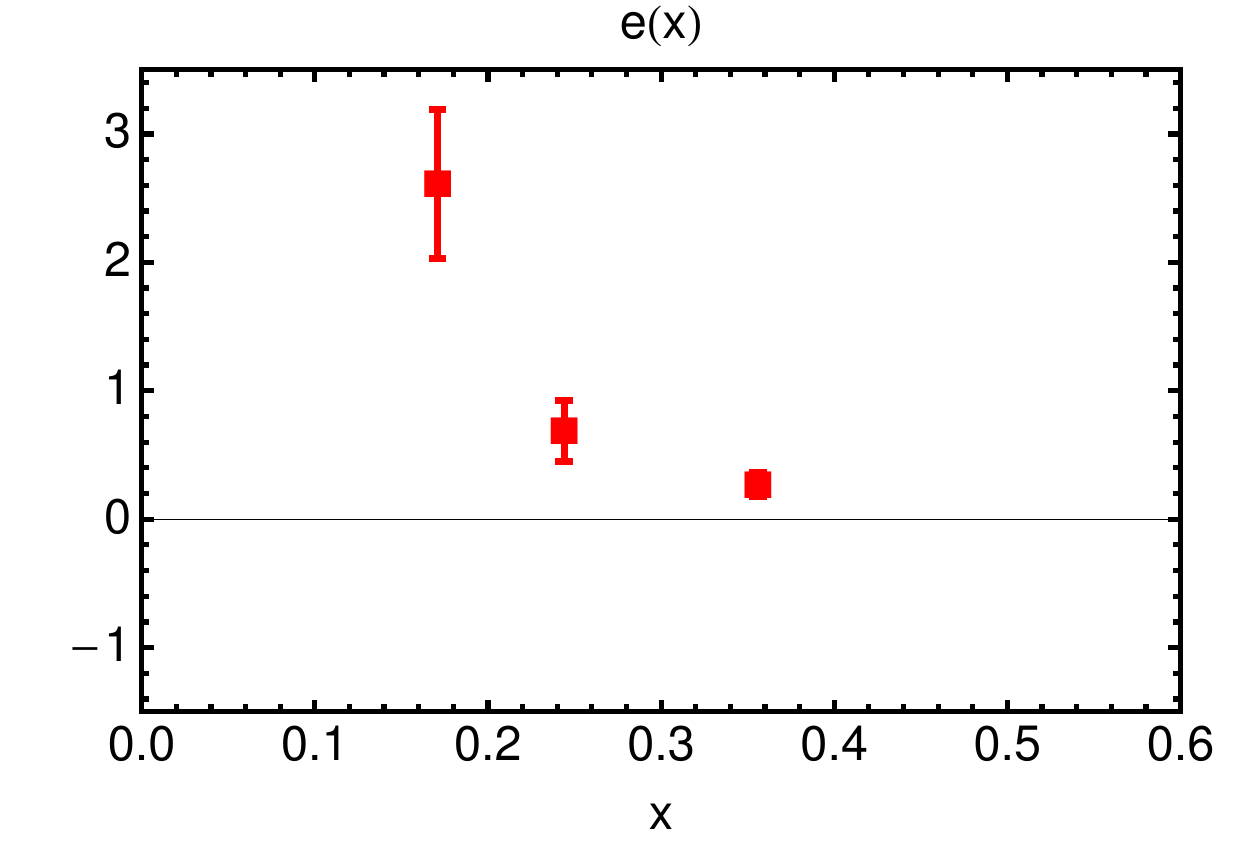} 
\caption{ The valence combination $e(x)\equiv4 e^{u_V}(x_i, Q_i^2)/9-e^{d_V}(x_i, Q_i^2)/9$ (see text). The error bars correspond to the propagation of the experimental and DiFF errors. } 
\label{fig:eww}
\end{figure}

An extraction of the twist-3 PDF, $e(x)$, through the  analysis of the  preliminary data~\cite{Pisano:2014ila} for the $\sin\phi_R$-moment of the beam-spin asymmetry for  dihadron Semi-Inclusive DIS at CLAS at 6 GeV was proposed in Ref.~\cite{Courtoy:2014ixa}. It is illustrated in Fig.~\ref{fig:eww} in a scenario in which the asymmetry is dominated by the term containing the twist-3 PDF.  We find a non-zero $e(x)$ even relaxing that hypothesis. The statistics do not allow for a proper fit for now, what restricts the possibility of extracting the scalar charge accurately. 

Soon the  CLAS12 data~\cite{prop_ht_silvia}   will be analyzed and allow for more statistics and information on that particular function. The determination of the scalar charge will be a delicate task as the value of the function at $x=0$ is experimentally unaccessible. The solution will be to appeal to other properties of the subleading-twist function together with advanced fitting techniques.

The evaluation of the scalar charge represents a challenge for the hadronic physics community but an opportunity to experimentally explore the consequence of chiral symmetry on the one hand. 

%
%


\section{Hadron phenomenology and BSM}

The role of the scalar and tensor aspects of the nucleon becomes clear in beta decay where the hadronic characterization of the process is encoded through, among others, the scalar and tensor form factor ---which value at zero momentum transfer is nothing else than the corresponding charge. While there is no leptonic counterpart for both structures in the Standard Model, all quark bilinear Lorentz structures can be introduced through
an effective Lagrangian which is relevant for beta decay
observables. The beta decay observables, {\it e.g.} $b_0^+$ and the Fierz term $b$, involve products of the Beyond the Standard Model (BSM) couplings, $\epsilon_i$, and
the corresponding hadronic charges, $g_i$.
The scalar (S) and tensor (T)
operators, in particular, contribute linearly to the beta decay
parameters through their interference with the SM amplitude,
and they are, therefore, more easily detectable. 
We focus here on the following low-energy effective  interaction (see Ref.~\cite{Cirigliano:2013xha} for a review)
\begin{eqnarray}
\Delta{\cal L}_{\rm eff} 
&=&
- C_S \bar{p}n\, \bar{e}(\mathbb{I}-\gamma_5)\nu_e
- C_T \bar{p}\sigma_{\mu\nu}n\, \bar{e}\sigma^{\mu\nu}(\mathbb{I}-\gamma_5)\nu_e\quad,
\label{eq:leffq2} 
\end{eqnarray}
where $C_S =   G_F V_{ud} \sqrt{2} \epsilon_{S}  g_{S}$, and $C_T =   4 G_F V_{ud} \sqrt{2} \epsilon_{T}  g_{T}$ and $G_F\equiv \sqrt{2}g^2/(8 M_W^2)$ is the tree-level definition of the Fermi constant.
In  the SM, the $\epsilon_S$ and $\epsilon_T$ coefficients 
vanish leaving the well-known $(V-A)\times(V-A)$ structure generated by the exchange of a $W$ boson. 
\\

Precise measurements in beta decay set bounds on the scalar coupling and on the plane $\epsilon_S-\epsilon_T$ through, {\it i.e.}, $b_0^+$ from super-allowed Fermi transition as well as the Fierz term $b$. The latter can be expressed in terms of the factorized coupling~\cite{Bhattacharya:2011qm}
\begin{eqnarray}
\label{b:eq}
b& = &  \frac{2}{1+ 3\lambda^2} \left[ g_S \epsilon_S  - 12 g_T \epsilon_T \lambda \right]  < 10^{-3}\quad,
\end{eqnarray}
with $\lambda = g_A/g_V$. The precision limit on the Fierz term corresponds to future experiments expectations, so does 
the bound on $b_0^+$  given at $68\%$ CL in Ref.~\cite{Hardy:2008gy}:
\begin{eqnarray}
b_0^+&=& -2 g_S \epsilon_S < 0.0028(26)\quad.
\label{eqs:bb0}
\end{eqnarray}

To isolate the tensor coupling from Eq.~(\ref{b:eq}), in Refs.~\cite{Pattie:2013gka,Wauters:2013loa}, a global fit has been performed, establishing strong bounds on the combination $g_T\epsilon_T$:
\begin{eqnarray}
\left | g_T\epsilon_T \right | < 6\cdot 10^{-4}~~~~~~\mbox{(90\% CL)}~,
\label{eq:gtet}
\end{eqnarray}
which is expected to be improved by the next generation of experiments. 
\\

In order to extract a bound on the Wilson coefficients $\epsilon_{S,T}$ from Eqs.~(\ref{eqs:bb0}, \ref{eq:gtet}) it is necessary to know the value of the hadronic charge $g_{S,T}$. 
Hence, the sensitivity of beta decay measurements to an exotic scalar or tensor interaction depends on our knowledge of the scalar or tensor charge, respectively. 
The impact of data based determinations of the tensor charge has been studied in Ref.~\cite{Courtoy:2015haa}. Given the  sensitivity bounds on the Fierz term $b$, an error of about $10-15\%$ on the tensor charge is estimated to be sufficient. The values for $g_T$ mentioned above are still far from that precision. 
\\

\begin{figure}
\centering
\includegraphics[width=8cm]{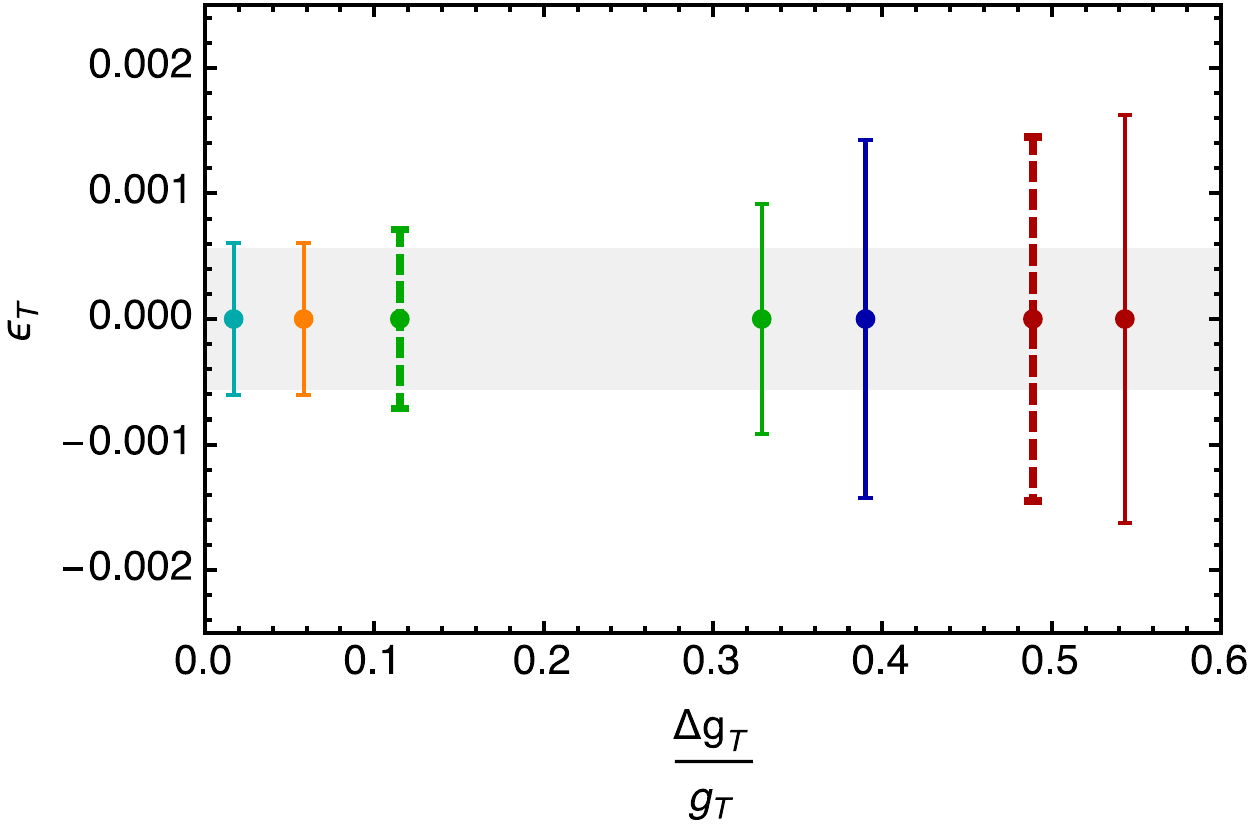} 
\caption{Bounds on $\epsilon_T$  obtained from precision measurements of beta decay using all current extractions and lattice QCD evaluations of the tensor charge $g_T$, plotted vs. the relative error, $\Delta g_T/g_T$: (turqoise)  Lattice QCD \cite{Green:2012ej,Bhattacharya:2013ehc}; (yellow) Lattice QCD \cite{Bali:2014nma};
(green) Deeply virtual $\pi^o$ and $\eta$ production \cite{Goldstein:2014aja}; (blue) single pion SIDIS \cite{Anselmino:2013vqa};  (red) dihadron SIDIS \cite{Radici:2015mwa}. The dashed lines are future projections based on JLab's upgrade.
All results were obtained using in the definition of $\Delta g_T/g_T$, each individual evaluation's $g_T$.  
The grey band gives the error assuming $\Delta g_T=0$, and the average $g_T$ of the three hadronic analyses considered in Ref.~\cite{Courtoy:2015haa}. The lattice evaluations from Refs. \cite{Green:2012ej,Bhattacharya:2013ehc} are indistinguishable.} 
\label{fig:tensorhadro}
\end{figure}

Originally, in Ref.~\cite{Courtoy:2015haa}, an alternative
to the standard Hessian evaluation of the errors was considered. The argument 
that 
both the lattice QCD and experimental extractions of the couplings from $\beta$-decay
are
affected by systematic or theoretical uncertainty invalidates the assumption of a gaussian distribution of the error around the central value. 
In order to deal with this situation we followed Ref.~\cite{Bhattacharya:2011qm} and we calculated the confidence interval on $\epsilon_T$ using the so-called R-Fit method~\cite{Hocker:2001xe}. In this scheme the theoretical likelihoods do not contribute to the $\chi^2$ of the fit and the corresponding QCD parameters take values within certain \lq\lq allowed ranges". In our case, this means that $g_T$ is restricted to remain inside a given interval, e.g. $0.37 \leq  g_T  \leq 1.25$  for the current determination from di-hadron SIDIS. 
The chisquare function is then given by
\begin{equation}
\chi^2 (\epsilon_T)  =  {\rm min}_{g_T} \left( \frac{\left[ g_T \epsilon_T \right]^{\rm exp}    - g_T \epsilon_T}{\delta \left[ g_T \epsilon_T \right]^{\rm exp}}\right)^2\quad,
\label{eq:chirfit}
\end{equation}
where the minimization is performed varying $g_T$ within its allowed range. In this approach, the bound on $\epsilon_T$ depends only on the lower limit of the tensor charge, as long as the experimental determination of $g_T\epsilon_T$ is compatible with zero at $1\sigma$.  In particular, the tensor charge given by Eq.~(\ref{eq:gtpavia})  leads to a bound $|\epsilon_T|<0.00162$. This bound is the largest found through hadron phenomenology (see Fig.~\ref{fig:tensorhadro}) but the error on $g_T$ is expected to decrease of about $10\%$ with the new JLab data.
For comparison, the bound obtained from the analysis of LHC data carried out in Ref.~\cite{Gonzalez-Alonso:2013uqa} is $|\epsilon_T|<0.0013$.

This error analysis presents shortcomings, {\it e.g.} the fact that Eq.~(\ref{eq:chirfit}) only depends on the lower bound on the tensor charge. The principal criticism would be that the R-fit method is a global method for data fitting, used here for the determination of a single parameter based on a fit output, Eq.~(\ref{eq:gtet}). Alternative error treatments will be consider in an extensive study of the hadronic structure impact on physics BSM.

On the other hand, a fully 2-dimensional analysis of the $\epsilon_S-\epsilon_T$ plane will be possible with more data for the scalar charge related observables.

\section{Conclusions}

\vspace{.2cm}

The phenomenology of Dihadron Fragmentation Functions integrates the idea that more structures and information become accessible when incorporating a dependence on the transverse momentum. The elegance of DiFF results in  less intricated and more easily taken care of observables, namely collinear objects and simple products, as opposed to convolutions. This particular feature allows for more flexibility in their statistical analysis. The parameterization of the third leading-twist PDF, {\it  transversity}, marked the first stage of an alternative approach to transverse momentum dependent functions. Then followed the tensor charge determination, made possible thanks to the flavor structure of the analyzed observables. Soon, data for subleading processes involving DiFF will be released, for which we will be able to consider the study of twist-3 PDFs. 

The latter  encode invaluable information on the low energy dynamics of quarks and gluons, so far hardly explored. The nonperturbative nature of nucleons influences numerous observables that involve them. The possibility of obtaining the scalar and tensor charges directly from experiment with sufficient precision, 
not having to deal with purely theoretical uncertainties of these quantities,
 gives a different weight to searches for BSM involving nucleons. 
 
 The results presented here concerning the determination of the scalar and tensor charges through Dihadron Fragmentation Functions can be complemented, in the tensor case, by the Deeply Virtual Meson Production analysis~\cite{Goldstein:2014aja} and the Single-hadron SIDIS approach~\cite{Ye:2016prn}.

\end{document}